\documentclass[11pt,a4paper]{article}
\usepackage{amsmath}
\usepackage{psfig}
\psfigurepath{./PS}
\usepackage{graphics}

\usepackage{graphicx}
\newcommand{\eps}{\epsilon}
\newcommand{\bitem}{\begin{itemize}}
\newcommand{\eitem}{\end{itemize}}
\newcommand{\goto}{\rightarrow}

\newcommand{\mmax}{\mathrm{max}}

\newcommand{\beqn}{\begin{equation}}
\newcommand{\eeqn}{\end{equation}}
\newcommand{\balign}{\begin{align}}
\newcommand{\ealign}{\end{align}}
\newcommand{\lam}{\lambda}
\usepackage{amssymb}
\begin{document}
% \nocite{*}
% \bibliography{apply}

% \title{A new estimator of non-Gaussian signatures: Application to CMB analysis}
\title{Cosmological non-Gaussian Signature Detection: \\
Comparing Performance of Different Statistical Tests}
\author{J. Jin \\
 Statistic Department, Purdue University,   \\
 150 N. University Street,  
 West Lafayette, IN 47907 USA \\ [12pt]
     J.-L. Starck\\
    DAPNIA/SEDI-SAP, Service d'Astrophysique, CEA-Saclay, \\
 F-91191 Gif-sur-Yvette Cedex, France \\  [12pt]
     D.L. Donoho  \\
    Department of Statistics, Stanford University,\\
Sequoia Hall, Stanford, CA 94305 USA\\   [12pt]
     N. Aghanim \\
   IAS-CNRS, Universit\'e Paris Sud, B\^atiment 121,\\
    F-91405, Orsay Cedex, France \\
Division of Theoretical Astronomy, 
 National Astronomical Observatory of
Japan,  \\
Osawa 2-21-1, Mitaka, Tokyo 181-8588, Japan \\ [12pt]
       O. Forni \\
   IAS-CNRS, Universit\'e Paris Sud, B\^atiment 121,\\
    F-91405, Orsay Cedex, France   
     }
\maketitle

\begin{abstract}
% Rappeler ce qu'est le fond cosmologique, qu'il est gaussien mais que des 
% composantes non-Gaussiennes peuvent etre la.
Currently, it appears that the best method for
non-Gaussianity detection in the Cosmic Microwave Background (CMB)
consists in calculating the kurtosis of the wavelet coefficients.  We
know that wavelet-kurtosis outperforms other methods such  as the
bispectrum, the genus, ridgelet-kurtosis and curvelet-kurtosis on an empirical basis,  but
relatively few studies have compared other transform-based
statistics, such as extreme values, or more recent tools such as Higher Criticism (HC), 
 or proposed `best possible' choices for such statistics. 

In this paper we consider two models for transform-domain coefficients: (a) a power-law model, 
which seems suited to the wavelet coefficients of simulated cosmic strings; and (b) a sparse mixture model, 
which seems suitable for the curvelet coefficients of filamentary structure.  
For model (a), if power-law behavior holds with finite $8$-th moment,   
excess kurtosis is an asymptotically optimal detector, but if the $8$-th moment is not finite, 
a test based on extreme values is asymptotically optimal. 
For model (b), if the transform coefficients are very sparse, a recent test, 
Higher Criticism, is an optimal detector, but if they are dense, kurtosis is an optimal detector. 
Empirical wavelet coefficients 
of simulated cosmic strings have power-law character, infinite $8$-th moment, 
while curvelet coefficients of the simulated cosmic strings are not very sparse.  
In all cases, excess kurtosis seems to be an effective test in moderate-resolution  imagery.
\end{abstract}

\section{Introduction}

The Cosmic Microwave Background (CMB), discovered in 1965 by Penzias
and Wilson \cite{gauss:penzias65}, is a relic of radiation emitted some
13 billion years ago, when the Universe was about 370.000 years
old. This radiation exhibits characteristic of  an almost 
perfect blackbody at a temperature of 2.726 Kelvin as measured by the
FIRAS experiment on board COBE satellite \cite{gauss:fixsen96}. The DMR
experiment, again on board COBE, detected and measured angular small
fluctuations of this temperature, at the level of a few tens of micro
Kelvin, and at angular scale of about 10 degrees \cite{gauss:smoot92}.
These so-called temperature anisotropies were predicted as the
imprints of the initial density perturbations which gave rise to
present large scale structures as galaxies and clusters of galaxies.
This relation between the  present-day universe and its initial conditions
has made the CMB radiation one of the preferred tools of cosmologists
to understand the history of the universe, the formation and evolution
of the cosmic structures and physical processes responsible for them
and for their clustering.

As a consequence, the last several years have been a particularly
exciting period for observational cosmology focussing on  the
CMB. With CMB balloon-borne and ground-based experiments such as TOCO
\cite{gauss:miller99}, BOOMERanG \cite{gauss:bernardis00}, MAXIMA
\cite{gauss:hanany00}, DASI \cite{gauss:halverson02} and Archeops  \cite{gauss:benoit03}, 
a firm detection of the so-called ``first peak'' in the
CMB anisotropy angular power spectrum at the degree scale was
obtained. This detection was very recently confirmed by the WMAP
satellite \cite{gauss:bennett03},  which detected also the second and
third peaks. WMAP satellite mapped the CMB temperature fluctuations
with a resolution better that 15 arc-minutes and a very good accuracy
marking the starting point of a new era of precision cosmology that
enables us to use the CMB anisotropy measurements to constrain the
cosmological parameters and the underlying theoretical models.

\begin{figure}[htb]
\centering
\caption{Courtesy of the WMAP team (reference to the website).
All sky map of the CMB anisotropies measured by the WMAP satellite.}
\label{fig_wmap}
\end{figure}
%\begin{figure}[htb]
%\centerline{
%\hbox{
%\psfig{figure=wmapcmb.ps,bbllx=0.cm,bblly=0cm,bburx=21.5cm,bbury=11.cm,height=10.5cm,width=16cm,clip=} }}
%\caption{Courtesy of the WMAP team (reference to the website).
%All sky map of the CMB anisotropies measured by the WMAP satellite.}
%\label{fig_wmap}
%\end{figure}

In the framework of adiabatic cold dark matter models, the position,
amplitude and width of the first peak indeed provide strong evidence
for the inflationary predictions of a flat universe and a
scale-invariant primordial spectrum for the density perturbations.
Furthermore, the presence of second and third peaks, confirm the
theoretical prediction of acoustic oscillations in the primeval plasma
and shed new light on various cosmological and inflationary
parameters, in particular, the baryonic content of the universe.  The
accurate measurements of both the temperature anisotropies and
polarised emission of the CMB will enable us in the very near future
to break some of the degeneracies that are still affecting parameter
estimation. It will also allow us to probe more directly the
inflationary paradigm favored by the present observations. 

Testing the inflationary paradigm can also be achieved through 
detailed study of the statistical nature of the CMB anisotropy distribution.  In
the simplest inflation models, the distribution of CMB temperature
fluctuations should be Gaussian,  and this Gaussian  field is completely determined
by its power spectrum. However,  many models such as  multi-field inflation
(e.g. \cite{bernardeau2002} and references therein), super strings or
topological defects, predict non-Gaussian contributions to the initial
fluctuations \cite{gauss:luo94,gauss:jaffe94,gauss:gangui94}. The statistical properties 
of the CMB should discriminate models of  the early universe.
Nevertheless, secondary effects like the inverse Compton scattering,
the Doppler effect, lensing and others add their own contributions to
the total non-Gaussianity.

All these sources of non-Gaussian signatures might have different
origins and thus different statistical and morphological
characteristics.  It is therefore not surprising that a large number
of studies have recently been devoted to the subject of the detection
of non-Gaussian signatures. Many approaches have been investigated:  Minkowski functionals and the morphological statistics
\cite{gauss:novikov00,gauss:shandarin02}, the bispectrum (3-point
estimator in the Fourier domain)
\cite{gauss:bromley99,gauss:verde00,gauss:phillips01}, the trispectrum
(4-point estimator in the Fourier domain) \cite{gauss:kunz01}, wavelet
transforms
\cite{gauss:aghanim99,gauss:forni99,gauss:hobson99,gauss:barreiro01,gauss:cayon01,gauss:jewell01,starck:sta03_1},
and the curvelet transform \cite{starck:sta03_1}. Different wavelet methods have been studied,
such as the isotropic \`a trous algorithm \cite{starck:book98} and the
bi-orthogonal wavelet transform \cite{ima:mallat98}. (The bi-orthogonal wavelet transform 
was found to be the most sensitive to  non-Gaussianity
\cite{starck:sta03_1}).  In
\cite{gauss:aghanim03,starck:sta03_1}, it was shown that the wavelet
transform was a very powerful tool to detect the non-Gaussian
signatures. Indeed, the excess kurtosis (4th moment) of the wavelet
coefficients  outperformed all the other methods (when the signal is characterised by a non-zero 4th moment).

Nevertheless, a major issue of the non-Gaussian studies in CMB remains
our ability to disentangle all the sources of non-Gaussianity from one
another. Recent progress has been made on the discrimination between
different possible origins of non-Gaussianity. Namely, it was
possible to separate the non-Gaussian signatures associated with
topological defects (cosmic strings (CS)) from those due to Doppler effect
of moving clusters of galaxies (both dominated by a Gaussian CMB
field) by combining the excess kurtosis derived from both the wavelet
and the curvelet transforms \cite{starck:sta03_1}.

This success argues for  us to construct
a ``toolkit'' of well-understood and sensitive methods for
probing different aspects of the non-Gaussian signatures.

In that spirit, the goal of the present study is to
consider the advantages and limitations of detectors which
apply kurtosis to transform coefficients of image data.
We will study plausible models for transform
coefficients of image data and compare the
performance of tests based on
kurtosis of transform coefficients to
other types of statistical diagnostics.

At the center of our analysis are two facts

[A] The wavelet/curvelet coefficients of CMB are Gaussian (we implicitly
assume the most simple inflationary scenario).

[B] The wavelet/curvelet coefficients of topological defect and Doppler
effect simulations are non-Gaussian.

We develop tests for non-Gaussianity for
two models of statistical behavior of transform coefficients.
The first, better suited for wavelet
analysis, models transform coefficients
of cosmic strings as following a power law.
The second, theoretically better suited for
curvelet coefficients, assumes that the
salient features of interest are actually filamentary
(it can be residual strips due do a non perfect calibration),
which gives the curvelet coefficients a sparse
structure.  

We review some  basic ideas
from detection theory, such as likelihood ratio detectors, and explain why
we prefer non-parametric detectors, valid across
a broad range of assumptions.

In the power-law setting, we  consider two
kinds of non-parametric detectors.
The first, based on kurtosis, is asymptotically
optimal in the class of weakly dependent symmetric
non-Gaussian contamination with finite 8-th moments.
The second, the Max, is shown to be asymptotically optimal in the class of weakly dependent symmetric non-Gaussian 
contamination with infinite 8-th moment.
While the evidence seems to be that wavelet coefficients of CS
have about 6 existing moments -- indicating
a decisive advantage for extreme-value statistics --
the performance of kurtosis-based
tests and Max-based tests on moderate sample sizes (eg. 64K transform
coefficients) does not follow the asymptotic theory;
excess kurtosis works better at these sample sizes.

In the sparse-coefficients setting, we consider
kurtosis, the Max, and a recent statistic called
{\em Higher Criticism} (HC) \cite{gauss:lin02}.
Theoretical analysis suggests that
curvelet coefficients of filamentary features
should be sparse, with about $n^{1/4}$ substantial
nonzero coefficients out of $n$ coefficients in
a subband; this level of sparsity would argue in
favor of Max/HC. However, empirically,
the curvelet coefficients of actual CS simulations
are not very sparse. It turns out that 
kurtosis outperforms  Max/HC in simulation.

Summarizing, the  work reported here seems to show that for
all transforms considered, the excess kurtosis outperforms
alternative methods despite their strong theoretical motivation.
A reanalysis of the theory supporting those methods shows that the
case for kurtosis can also be justified theoretically based on observed
statistical properties of the transform coefficients not used in the original
theoretic analysis.

\section{Detecting Faint Non-Gaussian Signals Superposed on a Gaussian Signal}
\label{sec:Theory}
The superposition  of a non-Gaussian signal with a Gaussian signal can be modeled as  $Y =  N  + G$,  where $Y$ is the observed image, $N$ is the non-Gaussian component and $G$ is the Gaussian component.  
We are interested in using transform coefficients to test whether $N \equiv 0$  or not.

\subsection{Hypothesis Testing and Likelihood Ratio Test (LRT).}  \label{subsec:LRT}
Transform coefficients of various kinds  [Fourier, wavelet, etc.] have been used 
for detecting non-Gaussian behavior in numerous studies.     Let $X_1, X_2, \ldots, X_n$ be the transform coefficients of $Y$; we model these as  
\begin{equation}    \label{EqAlt}
X_i = \sqrt{1 - \lam} \cdot  z_i + \sqrt{\lam} \cdot w_i,  \qquad 0< \lam < 1, 
\end{equation}
where  $\lam >0$ is a parameter, $z_i \stackrel{iid}{\sim} N(0,1)$ are  the transform coefficients   of the Gaussian component $G$,  $w_i \stackrel{iid}{\sim} W$  are the transform coefficients of the non-Gaussian component  $N$, and $W$ is  some unknown  symmetrical distribution.  Here without loss of generality, we  assume the standard deviation for both  $z_i$ and $w_i$ are  $1$.

Phrased in statistical terms,   the problem of detecting  the existence of a non-Gaussian component  is equivalent  to discriminating between the hypotheses:  \begin{align}
&H_0: \;\;\;   X_i = z_i,  \label{EqHypo1}   \\
&H_1:   X_i = \sqrt{1 - \lam } \cdot z_i  + \sqrt{\lam} \cdot  w_i,   \qquad 0 < \lam < 1,   \label{EqHypo2}
\end{align}
and  $N \equiv 0$ is equivalent to $\lam \equiv 0$.  
We call  $H_0$ the {\it null hypothesis $H_0$}, and $H_1$ the {\it alternative hypothesis}.

\begin{figure}
\centering
\caption{Detectable regions  in the $\alpha-r$ plane.  With $(\alpha,r)$ in the white region on the top or the undetectable region, all methods completely fail for detection. With $(\alpha,r)$ in the white region on the bottom,  both excess kurtosis and Max/HC are able to detect reliably.      While in the blue region to the left,  Max/HC is able to detect reliably, but excess kurtosis completely fails, and in the yellow region to the right, excess kurtosis is able to detect reliably, but Max/HC  completely fail.     }
\label{Figure:Detect}
\end{figure}
%\begin{figure}[htb]
%\centerline{
%\hbox{
%\psfig{figure=CSDetectBoundary.eps,bbllx=1.5cm,bblly=8.cm,bburx=19.5cm,b
%bury=23cm,height=6cm,width=7.5cm,clip=}
%\psfig{figure=CSDetectRegions.eps,bbllx=1.5cm,bblly=8.cm,bburx=19..5cm,bbury=23cm,height=6cm,width=7.5cm,clip=}
% \psfig{figure=CSDetectBoundarye.eps,height=5cm,width=6cm,clip=}
% \psfig{figure=CSDetectRegion.eps,height=5cm,width=6cm,clip=}
%}}
% \includegraphics[height = 4 in]{Picture/CSDetectBoundary.pdf}
%\includegraphics[height = 4 in]{Picture/CSDetectRegionCopy.pdf}
%\caption{Left panel: Detection Boundary in the $\alpha-r$ plane. 
%The solid curve is the detection boundary of LRT, above which is not possible 
%to detect, and below which it is possible to reliably detect, the  
%dotted line segment and solid line segment together is the detection 
%boundary for Kurtosis,  the dotted curve and the solid curve together 
%is the detection boundary of Max/HC.  Right panel: detectable regions for %Kurtosis, Max/HC.}
%\label{Figure:Detect}
%\end{figure}
When both $W$ and $\lam$ are known,  then the optimal test for Problem (\ref{EqHypo1}) - (\ref{EqHypo2}) is simply the Neyman-Pearson  Likelihood ratio test (LRT), \cite[Page 74 ]{Lehmann}. The size of $\lam = \lam_n$  for which reliable 
discrimination between $H_0$ and $H_1$ is possible can be derived using asymptotics.      If  we assume that the tail probability of $W$ decays algebraically, 
\begin{equation} \label{EqDefineAlg}
\lim_{x \goto \infty}   x^{\alpha}  P\{|W| > x\}  = C_{\alpha},  \qquad \mbox{$C_{\alpha}$ is a constant},
\end{equation}
(we say $W$ has a power-law tail),  and we calibrate $\lam$ to decay with $n$, so that increasing amounts of data are offset by increasingly hard challenges: 
\begin{equation}   \label{EqDefineLam}
\lam = \lam_n  = n^{-r},  
\end{equation}
then there is a {\it threshold effect} for the detection problem (\ref{EqHypo1}) - (\ref{EqHypo2}).  In fact,  define:
\begin{equation} \label{EqDetectBoundary}
\rho^*_1(\alpha) = 
\left\{ \begin{array}{ll}
2/\alpha, &\   \  \alpha \leq 8,\\
1/4, &\     \       \alpha > 8,
\end{array}
\right.
\end{equation}
then as $n \goto \infty$,   LRT is able to reliably detect  for large $n$
when  $r < \rho^*_1(\alpha)$, and is unable  to detect when $r >  \rho^*_1(\alpha)$; this is proved in \cite{DJ04b}.  
Since LRT is optimal, 
 it is not possible for any statistic to  reliably detect when $r >  \rho^*_1(\alpha)$.  
We call the curve $r = \rho^*_1(\alpha)$ in the $\alpha$-$r$ plane  the {\it detection  boundary}; see Figure \ref{Figure:Detect}.

In fact,  when $r  < 1/4$,  asymptotically LRT is  able to reliably detect whenever $W$ has a finite $8$-th moment,  even without the assumption that $W$ has a power-law tail. Of course, the case that $W$ has an infinite $8$-th moment is more complicated, but if   $W$ has a power-law tail, then LRT is also able to reliably detect if $r < 2/\alpha$. 

%One component of the above result is that, by assuming  $W$ has an %$\alpha$-algebraic tail with $\alpha > 8$, then when $r < \frac{1}{4}$,  LRT is %able to reliably detect;  and when $r > \frac{1}{4}$, no statistic is able to detect.  %It is interesting to notice here that, this part of the conclusion will still hold %when  
%the condition of requiring $W$ to have an algebraic tail is largely relaxed:  in %fact,  the same conclusion still holds if we only require $E[W^8] < \infty$.  It is %interesting to notice here that,  when $W$ has an $\alpha$-algebraic tail, %$E[W^8] < \infty$ if and only if $\alpha > 8$.

Despite its optimality, LRT is not a practical procedure. To apply LRT, one needs to  specify  the value of  $\lam$ and  the distribution of  $W$,  which seems unlikely to be available. We need   non-parametric detectors,  which can be implemented without any knowledge of  $\lam$ or $W$,  and depend on $X_i$'s only.  In the section below, we are going to introduce two non-parametric detectors:  excess kurtosis and Max; later in Section \ref{sec:HC}, we will introduce a third non-parametric detector:  Higher Criticism (HC).

\subsection{Excess Kurtosis and Max}    \label{subsec:Statistics}
We pause to  review the concept of $p$-value briefly.  
For a statistic $T_n$,  the $p$-value is the probability of seeing  equally extreme results under the  null  hypothesis:
\[
p = P_{H_0} \{ T_n  \geq t_n(X_1,X_2, \ldots,X_n) \}; 
\] 
here $P_{H_0}$ refers to probability under $H_0$, and $t_n(X_1,X_2, \ldots,X_n)$ is the observed value of statistic $T_n$.
Notice that the smaller the $p$-value, the stronger the evidence against the null hypothesis. A natural decision rule based on $p$-values rejects  the null when $p <  \alpha$ for some selected level $\alpha$, and a convenient choice is  $\alpha = 5\%$.  When the null hypothesis is indeed true,  the $p$-values for any statistic  are distributed as  uniform $U(0,1)$. This implies that  the  $p$-values  provide a common scale for comparing  different statistics. 

We now introduce  two statistics for comparison. 

{\bf Excess Kurtosis ($\kappa_n$)}.  Excess kurtosis is a widely used statistic, based on  the $4$-th moment. 
For any (symmetrical) random variable $X$, the kurtosis is:
\[
\kappa(X) = \frac{EX^4}{(EX^2)^2} -3. 
\]
The kurtosis measures a kind  of  departure of $X$  from  Gaussianity,  as $\kappa(z) =  0$.  

Empirically,  given  $n$ realizations of $X$,  the excess kurtosis statistic is defined as: 
\begin{equation}  \label{EqDefineK}
\kappa_n(X_1, X_2,\ldots,X_n)  = \sqrt{\frac{n}{24}} \biggl[ \frac{\frac{1}{n}\sum_i  X_i^4}{(\frac{1}{n}  \sum_i X_i^2)^2}  - 3  \biggr].
\end{equation} 
When the null is true, the excess kurtosis statistic is asymptotically normal:
\[
\kappa_n(X_1, X_2,\ldots,X_n)  \rightarrow_{w}  N(0,1), \qquad n \goto \infty,
\]
thus for large $n$, the $p$-value of the excess kurtosis  is  approximately:
\[
\tilde{p} = \bar{\Phi}^{-1} (\kappa_n(X_1, X_2,\ldots,X_n)), 
\]
where $\bar{\Phi}(\cdot)$ is the  survival function (upper tail probability)  of $N(0,1)$. 

It is proved in \cite{DJ04b} that  the excess kurtosis is asymptotically  optimal for the hypothesis testing of (\ref{EqHypo1}) -  (\ref{EqHypo2}) if 
\[
E [W^8] < \infty.
\]
However, when $E[W^8] = \infty$,  even though  kurtosis  is well-defined ($E[W^4] < \infty$), 
there are situations in which LRT  is able to reliably detect but  excess kurtosis  completely fails.
In fact, by assuming  (\ref{EqDefineAlg}) - (\ref{EqDefineLam}) with an $\alpha < 8$,  
if $(\alpha,r)$ falls into the blue region of Figure~\ref{Figure:Detect},  then LRT is able 
to reliably detect, however, excess kurtosis completely fails. This shows that in such cases,  
excess kurtosis is not optimal;  see \cite{DJ04b}. 

{\bf Max ($M_n$).}   The largest (absolute) observation is a classical and frequently-used 
non-parametric  statistic:
\[
M_n =  \mmax(|X_1|,|X_2|,\ldots, |X_n|),
\] 
under the null hypothesis, 
\[
M_n  \approx \sqrt{2 \log n},
\]
and moreover,  by normalizing  $M_n$ with constants $c_n$ and $d_n$, the resulting statistic 
converges to the Gumbel distribution $E_v$,   whose cdf is $e^{-e^{-x}}$:
\[
\frac{M_n - c_n}{d_n}  \rightarrow_{w}    E_v,
\]
where approximately
\[
d_n = \frac{\sqrt{6} S_n}{\pi}, \qquad  c_n = \bar{X} - 0.5772 d_n; 
\]
here $\bar{X}$ and $S_n$ are the sample mean and sample  standard deviation of $\{X_i\}_{i=1}^n$ respectively.  
Thus a good approximation of the  $p$-value for $M_n$ is:
\[
\tilde{p} =  \mathrm{exp}(-\mathrm{exp}(-\frac{M_n - c_n}{d_n})). 
\]
We have tried the above experiment for $n = 244^2$, and found that taking 
$c_n = 4.2627$, $d_n = 0.2125$ gives  a  good approximation.  

Assuming  (\ref{EqDefineAlg}) - (\ref{EqDefineLam})  and $\alpha < 8$, or  $\lam = n^{-r}$ and  
that $W$ has a power-law tail with $\alpha <  8$,  it is proved in \cite{DJ04b} that Max  is optimal for hypothesis testing (\ref{EqHypo1}) - (\ref{EqHypo2}). 
Recall if we further assume  $\frac{1}{4} < r < \frac{2}{\alpha}$, then asymptotically, 
excess kurtosis completely fails; however, Max is able to reliably detect and is competitive to LRT. 

On the other hand,  recall that excess kurtosis is optimal for  the case $\alpha > 8$. 
In comparison, in this case,  Max is not optimal. In fact, if we further 
assume $ \frac{2}{\alpha} < r < \frac{1}{4}$, then  excess kurtosis is able to reliably detect,  but
Max will completely fail. 

In Figure \ref{Figure:Detect}, we compared the detectable regions of the  excess 
kurtosis and Max in  the $\alpha$-$r$ plane. 

To conclude this section, we mention an alternative way to approximate 
the $p$-values for any statistic $T_n$.  This alternative way is important in 
case that an asymptotic (theoretic) approximation is poor for moderate large $n$, 
an example is the  statistic  $HC_n^*$ we will introduce in Section \ref{sec:HC};     
this alternative way  is helpful even when the asymptotic approximation  is accurate.   
Now the idea is,  under the null  hypothesis,  we simulate a large number ($N = 10^4$ or more) 
of $T_n$: $T_n^{(1)}, T_n^{(2)}, \ldots, T_n^{(N)}$, we then tabulate them. For the observed 
value $t_n(X_1, X_2, \ldots, X_n)$,  the $p$-value will then be well approximated by:
\[
\frac{1}{N} \cdot \#\{k:  \;  T_n^{(k)}  \geq  t_n(X_1, X_2, \ldots, X_n)\},
\]
and the larger the $N$,  the better the approximation.

\subsection{Heuristic Approach}  \label{subsec:Intuition}
We have exhibited a phase-change phenomenon, where the asymptotically optimal test changes depending on power-law index $\alpha$. 
In this section, we develop
a heuristic analysis of detectability and  phase change. 

The detection property  of Max follows from comparing the ranges of data. 
Recall that $X_i = \sqrt{1 -\lam_n} \cdot z_i  + \sqrt{\lam_n} \cdot w_i$,   
the range of  $\{z_i\}_{i=1}^n$  is roughly
$(-\sqrt{2 \log n} , \sqrt{2 \log n})$, 
and the range of  $\{\sqrt{\lam_n} \cdot w_i\}_{i=1}^n $  is 
$
\sqrt{\lam_n} \cdot (- n^{\frac{1}{\alpha}},    n^{\frac{1}{\alpha}})  
=  (- n^{\frac{1}{\alpha} - \frac{r}{2}},    n^{\frac{1}{\alpha}  - \frac{r}{2} })$; 
so heuristically,
\[
M_n  \approx  \mmax\{ \sqrt{2 \log n},   n^{\frac{1}{\alpha} - \frac{r}{2}} \};
\]
for  large $n$, notice that:
\[
n^{\frac{1}{\alpha} - \frac{r}{2}}  \gg   \sqrt{2 \log n},  \;\; \mbox{if $r < \frac{2}{\alpha}$},  
\qquad  \qquad  n^{\frac{1}{\alpha} - \frac{r}{2}}  \ll   \sqrt{2 \log n},  \;\; \mbox{if  $r > \frac{2}{\alpha}$}, 
\]
thus if and only if $r < \frac{2}{\alpha}$,  $M_n$ for the alternative will differ significantly   
from $M_n$ for the null, and so the criterion for detectability by Max is $r < \frac{2}{\alpha}$. 

Now we study   detection by   excess kurtosis. Heuristically, 
\[
\kappa_n \approx (1/\sqrt{24})  \cdot \kappa( \sqrt{1 -\lam_n}  \cdot z_i  + \sqrt{\lam_n} \cdot w_i)  = (1/\sqrt{24}) \cdot  \sqrt{n} \cdot  \lam_n^2  \cdot  \kappa(W) =  O(n^{1/2 - 2r}),
\]
thus  if and only if  $r < \frac{1}{4}$ will  $\kappa_n$ for the alternative differ  
significantly  from $\kappa_n$  under  the null, and so the criterion for detectability 
by excess kurtosis is $r < \frac{1}{4}$.

\begin{figure}
\centering
\caption{Primary Cosmic Microwave Background anisotropies (left) and simulated  cosmic string map  (right).}
\label{fig_cmb}
\end{figure}
%\begin{figure*}[htb]
%\centerline{
%\hbox{
%\psfig{figure=fig_cmb.ps,bbllx=1.9cm,bblly=12.7cm,bburx=14.6cm,bbury=25.4%cm,height=7cm,width=7cm,clip=}
%\psfig{figure=fig_cs.ps,bbllx=1.9cm,bblly=12.7cm,bburx=14.6cm,bbury=25.4c%m,height=7cm,width=7cm,clip=}
%}}
%\caption{Primary Cosmic Microwave Background anisotropies (left) and 
% cosmic string simulated map (right).}
%\label{fig_cmb}
%\end{figure*}

This analysis shows the reason for  the phase change.   In Figure \ref{Figure:Detect}, when the parameter $(\alpha,r)$ is in the blue region, for sufficiently large $n$, $n^{\frac{1}{\alpha} - \frac{r}{2}}  \gg   \sqrt{2 \log n}$ and the strongest evidence against the null is in the tails of the data set, which $M_n$ is indeed using. However, when   $(\alpha,r)$
moves from the blue region to the yellow region, $n^{\frac{1}{\alpha} - \frac{r}{2}}  \ll   \sqrt{2 \log n}$, the tails no longer contain any important evidence against the null,  instead, the central part of the data set  contain the  evidence.   By symmetry, the $1^{st}$ and the $3^{rd}$ moments vanishes, and the $2^{nd}$ moment is $1$ by the normalization; so the excess kurtosis is in fact the most promising candidate of detectors based on moments.

The heuristic analysis is the essence for theoretic proof as well as 
empirical experiment. Later in Section \ref{sec:Simulation}, we will 
have more discussions for comparing the excess kurtosis with Max down this vein.  

\section{Wavelet Coefficients of Cosmic Strings}   
\label{sec:AlgTail}

\subsection{Simulated Astrophysical Signals}
The temperature anisotropies of the CMB contain the contributions of
both the primary cosmological signal, directly related to the initial
density perturbations, and the secondary anisotropies.  The latter are
generated after matter-radiation decoupling \cite{white2002}.  They
arise from the interaction of the CMB photons with the neutral or
ionised matter along their path
\cite{sunyaev80,ostriker86,vishniac87}.

In the present study, we assume that the primary CMB anisotropies are
dominated by the fluctuations generated in the simple single field
inflationary Cold Dark Matter model with a non-zero cosmological
constant. The CMB anisotropies have therefore a Gaussian distribution. We
allow for a contribution to the primary signal from topological
defects, namely cosmic strings (CS), as suggested in
\cite{gauss:bouchet00}. 

We use for our simulations the cosmological
parameters obtained from the WMAP satellite \cite{astro:bennett2003}
and a normalization parameter $\sigma_8=0.9$. Finally, we obtain 
the so-called ``simulated observed map'', $D$, that contains the two
previous astrophysical components. It is obtained from
$D_\lambda=\sqrt{1 - \lambda}\mathrm{CMB}+\sqrt{\lambda}\mathrm{CS}$,
where $\mathrm{CMB}$ and $\mathrm{CS}$ are
respectively the CMB and the cosmic string simulated
maps. $\lambda=0.18$ is an upper limit constant derived by
\cite{gauss:bouchet00}. All the simulated maps  
have $500 \times 500$ pixels with a resolution of
1.5 arcminute per pixel.

\begin{figure}
\centering
\caption{Simulated observation containing the CMB and the CS ($\lambda=0.18$).}
\label{fig_cmbcs}
\end{figure}

\subsection{Evidence for $E[W^8] = \infty$} 
 \begin{table}
$$
\begin{tabular}{|r|r|r|r|r|r|} 
\hline 
size of & $4$-th &$5$-th  & $6$-th &$7$-th  & $8$-th \\
sub-sample&  moment &   moment  &  moment &moment & moment \\
\hline 
$n$   &  $30.0826$  &  $262.6756$ &  $2.7390 \times 10^3$ & $3.2494 \times 10^4$ & $4.2430 \times 10^5$ \\
\hline
$n/2$   &  $29.7100$  & $256.3815$ & $2.6219 \times 10^3$ & $2.9697 \times 10^4$ & $3.7376 \times 10^5$ \\
\hline
$n/2^2$   &  $29.6708$  &  $250.0520$ &  $2.4333 \times 10^3$ &$2.6237 \times 10^4$ & $3.0239 \times 10^5$ \\
\hline
$n/2^3$   &  $29.4082$  &$246.3888$ &  $2.3158 \times 10^3$ &$2.4013 \times 10^4$ & $2.3956 \times 10^5$ \\
\hline
$n/2^4$   &  $27.8039$  &$221.9756$ &  $1.9615 \times 10^3$ & $1.9239 \times 10^4$ &$1.8785 \times 10^5$ \\
\hline
\end{tabular}
$$
\caption{Empirical estimate $4$-th, $5$-th, $6$-th, $7$-th, and $8$-th moments calculated  using  a  subsamples of size $n/2^k$ of $\{|w_i|\}_{i=1}^n$,  with $k = 0, 1, 2, 3, 4$. The table suggests that  the $4$-th, $5$-th, and $6$-th moments are finite, but the $7$-th and $8$-th moments are infinite.}
\label{table:moments}
\end{table}

For the wavelet coefficients on the finest scale of the cosmic string map in the right panel of Figure  \ref{fig_cmb},  
by throwing away all the coefficients related to pixels on the edge of the map, we have $n = 244^2$ coefficients; 
we then normalize these coefficients so that the empirical  mean and standard deviation are $0$ and $1$ respectively; 
we denote the resulting dataset by  $\{w_i\}_{i=1}^n$.

Assuming $\{w_i\}_{i=1}^n$  are independent samples from a distribution  $W$, 
we have seen in Section 2 that, whether excess kurtosis is better than Max  
depends on the finiteness of  $E[W^8]$. We now analyze $\{w_i\}_{i=1}^n$  to learn about  $E[W^8]$.

Let 
\[
\hat{m}_8^{(n)}  = \frac{1}{n}  \sum_{i = 1}^n w_i^8, 
\]
be the empirical $8$-th moment of $W$ using $n$ samples.  In theory, if $E[W^8] < \infty$,  then $\hat{m}_8^{(n)}  \goto E[W^8] $ as $n \goto \infty$.  So one way to see if $E[W^8]$ is finite is to observe how $\hat{m}_8^{(n)}$ changes with $n$.   

Technically, since we only have $n = 244^2$ samples,  we can compare 
\[
\hat{m}_8^{(n/2^k)}, \qquad k = 0, 1,2,3,4;  
\]
if these values  are roughly the same, then there is strong evidence for $E[W^8] <  \infty$; otherwise, if they increase with sample size, that is  evidence for $E[W^8] =   \infty$.  Here $m_8^{(n/2^k)}$ is an estimate of $E[W^8]$ using $n/2^k$ sub-samples of $\{w_i\}_{i=1}^n$.  

For $k = 1, 2,3,4$, to obtain  $\hat{m}_8^{(n/2^k)}$, we randomly draw sub-samples  of size $n/2^k$ from $\{w_i\}_{i=1}^n$, and then take the average of the $8$-th power of  this subsequence; we  repeat this process $50,000$ times, and we let  $\hat{m}_8^{(n/2^k)}$ be the median of these $50,000$ average values. Of course when $k = 0$,  $\hat{m}_8^{(n/2^k)}$ is obtained  from  all $n$ samples. 

The results correspond  to the first wavelet band
are summarized in Table~\ref{table:moments}.  
>From the table, we have seen that 
$\hat{m}_8^{(n)}$  is significantly larger than  
 $\hat{m}_8^{(n/8)}$ and $\hat{m}_8^{(n/16)}$; 
this supports  that $E[W^8] = \infty$. Similar results were obtained from
the other bands. In comparison, in Table~\ref{table:moments}, we also list the 
$4$-th, $5$-th, $6$-th, and $7$-th moments.  It seems that the $4$-th, $5$-th, and $6$-th moments are finite, but the $7$-th and $8$-th moments are infinite.

\subsection{Power-law Tail of $W$}  \label{sec:Power}
Typical models for heavy-tailed data include exponential tails and power-law tails.  
We now compare  such models to the data on wavelet coefficients for $W$;   the 
Gaussian model is also included as comparison. 

We sort  the $|w_i|$'s in descending order, $|w|_{(1)} > |w|_{(2)}  > \ldots > |w|_{(n)}$,  
and take the $50$ largest samples $|w|_{(1)} > |w|_{(2)}  > \ldots > |w|_{(50)}$.  
For a power-law tail with index $\alpha$, we expect that for some constant $C_{\alpha}$, 
\[
\log(\frac{i}{n}) \approx  \log(C_{\alpha})  - \alpha \log(|w|_{(i)}),  \qquad 1 \leq i \leq 50, 
\]
so there is  a strong linear relationship between $\log(\frac{i}{n})$ and $\log(|w|_{(i)})$.  
Similarly, for  the exponential model, we expect a strong linear relationship 
between $\log(\frac{i}{n})$ and $|w|_{(i)}$, and for the Gaussian model, we expect a strong 
linear relationship between $\log(\frac{i}{n})$ and $|w|^2_{(i)}$.

For each model, to measure whether the ``linearity" is sufficient to explain the 
relationship between $\log(\frac{i}{n})$ and $\log(|w|_{(i)})$ (or $|w|_{(i)}$,  or $|w|^2_{(i)}$), 
we introduce the following $z$-score:
\begin{equation}  \label{EqDefineZ}
Z_i  = \sqrt{n} \biggl[ \frac{\hat{p}_i - i/n}{i/n(1 - i/n)} \biggr], 
\end{equation}
where $\hat{p}_i$ is the linear fit using each of the three models.  
If the resulting $z$-scores is  random and have no specific trend, 
the model is  appropriate; otherwise the model may need improvement.

%\begin{figure}[htb]   
%\centerline{
%\hbox{
% \includegraphics[height = 5 in]{Picture/Tail50.pdf}
%\psfig{figure=Tail50.eps,bbllx=1.5cm,bblly=8.cm,bburx=19.5cm,bbury=23cm,h%eight=10cm,width=14cm,clip=}
%}}
%\caption{Left panel:  from top to bottom, 
%plots of log-probability $\log(i/n)$ versus 
%$\log(|w|_{(i)}$, $|w|_{(i)}$, and $|w|^2_{(i)}$ for $1 \leq i \leq 50$,
% which corresponding 
%to the algebraic/exponential/Gaussian model we introduced 
%in Section \ref{sec:AlgTail} and $w$ corresponds to the wavelet coefficients
%of the first scale (i.e. highest frequencies).  
%Right panel:  from top to bottom, normalized $z$-score as 
%defined in (\ref{EqDefineZ}) for 
%the algebraic/exponential/Gaussian models, where again for $1 \leq i \leq %50$.}
%\label{Fig:Tail50}
%\end{figure}

The results are summarized in Figure~\ref{Fig:Tail50}. The 
power-law tail model seems
the most appropriate: the relationship between $\log(\frac{i}{n})$ and $\log(|w|_{(i)})$  looks very 
close to linear, the $z$-score looks very small, and the range of $z$-scores much narrower than 
the other two.   For the exponential model,  the linearity is fine at the first glance, however, the $z$-score  is  
 {\it decreasing}  with  $i$, which  implies that the tail is heavier 
than estimated.  The Gaussian model fits  much  worse than exponential. To summarize, there 
is strong evidence that the tail  follows a power-law.

Now we estimate the  index  $\alpha$ for the power-law tail. 
A widely-used  method for estimating $\alpha$  is the Hills' estimator \cite{gauss:hill75}:
\[
\hat{\alpha}_H^{(l)}     =   \frac{l+1}{ \sum_{i = 1}^l i \log(\frac{|w|_{(i)}}{|w|_{(i+1)}})} 
\]
where $l$ is the number of (the largest) $|w|_{(i)}$  to include for estimation.  
In our situation, $l = 50$ and 
 \[
\hat{\alpha}  =   \hat{\alpha}_H^{(50)}  = 6.134;
 \]
we also found that the standard deviation of  this estimate  $\approx 0.9$. 
Table \ref{tab_cs_alpha} gives estimates of $\alpha$  for each band
of the wavelet transform.
This shows that $\alpha$ is likely to be only slightly less than $8$: this means the performance of 
excess kurtosis and Max might be very close empirically.

%%%%%%%%
%%%%%%%%
%%%%%%%%
\begin{figure}
\centering
\caption{Left panel:  from top to bottom, 
plots of log-probability $\log(i/n)$ versus 
$\log(|w|_{(i)}$, $|w|_{(i)}$, and $|w|^2_{(i)}$ for $1 \leq i \leq 50$,
corresponding 
to the power-law/exponential/Gaussian models we introduced 
in Section \ref{sec:AlgTail};  $w$  are  the wavelet coefficients
of the finest scale (i.e. highest frequencies).  
Right panel:  from top to bottom, normalized $z$-score as 
defined in (\ref{EqDefineZ}) for 
the power-law/exponential/Gaussian models, where again for $1 \leq i \leq 50$.}
\label{Fig:Tail50}
\end{figure}

%%%%%%%
%%%%%%%
%%%%%%%
\begin{figure}
\centering
\caption{Top row: left panel,  from top to bottom, the fraction of detection for the excess kurtosis, $HC^*$ and Max, the $x$-axis is the corresponding $\lam$;  right panel,  from top to bottom, the fraction of detection for kurtosis, $HC^+$ and Max.  Bottom row:  left panel,  for top to bottom, ROC curves for the excess kurtosis, $HC^*$, and Max; right panel,  ROC curves for the excess kurtosis, $HC^+$, and Max. }
\label{Figure:ROCHC}
\end{figure}

\subsection{Comparison  of  Excess Kurtosis vs. Max  with Simulation}  
\label{sec:Simulation}
To test the results in Section \ref{sec:Power}, we now perform a small simulation experiment. 
A complete cycle includes the following steps. 
($n = 244^2$  and $\{w_i\}_{i=1}^n$  are the same as in 
 Section \ref{sec:Power}). 
\begin{enumerate}
\item Let $\lam$ range from $0$ to $0.1$ with increment $0.0025$.
\item Draw $(z_1,z_2,\ldots,z_n)$  independently from $N(0,1)$ to represent the transform  coefficients for CMB.
\item For each $\lam$, let
\[
X_i = X_i^{(\lam)}  = \sqrt{1-\lam}  z_i  + \sqrt{\lam} w_i,  \qquad \lam = 0,0.0025, \ldots, 0.1
\]
represent the transform coefficients for CMB + CS.
\item Apply detectors $\kappa_n$, $M_n$ to the $X_i^{(\lam)}$'s;   and 
obtain the $p$-values.
\end{enumerate}
We  repeated  the step 3-4  independently $500$ times.

Based on these simulations, first,  we have estimated the probability  of 
 detection under various  $\lam$,  for each  detector:
\[
\mbox{Fraction of  detections}   = \frac{\mbox{number of cycles with a $p$-value $ \leq 0.05$}}{500}.  
\]
Results are summarized 
in  Figure \ref{Figure:ROCHC}.  

Second, we pick out those simulated values for $\lam = 0.05$ alone, and 
plot the ROC curves for each detector.    
The ROC curve is a standard way to evaluate detectors \cite{ROC};    the $x$-axis gives the fraction of false alarms (the fraction of detections  when the null is 
true (i.e. $\lam = 0$));  the $y$-axis gives  the corresponding 
fraction of true detections). Results are shown  in   Figure \ref{Figure:ROCHC}. 
The figure suggests that  the excess kurtosis is slightly better than $M_n$. 
We also show an adaptive test, $HC_n$ in two forms ($HC_n^*$ and $HC_n^+$);   these will be described later. 

%%%%%%%%%
%%%%%%%%%
%%%%%%%%%
\begin{figure}
\centering
\caption{The $M$-$\kappa$ plane and the curve $\kappa = \kappa_0(M)$, 
where $M$ is the largest (absolute) observation  
of  $w_i$'s, and $\kappa$ is the empirical excess kurtosis of $w_i$'s, 
where $w_i$'s are  the wavelet coefficients of the simulated cosmic string.   Heuristically,  if $(M,\kappa)$ falls above the curve,   excess kurtosis will perform  better than Max.  The red star represent the points  of $(M,\kappa) = (17.48, 27.08)$ for the current data set $w_i$'s, which  is far above the curve.}
\label{Figure:Region}
\end{figure}

We now interpret. 
 As our analysis predicts  that $W$  has a power-law 
 tail  with $E[W^8] = \infty$,  it is surprising 
 that excess kurtosis still performs better than Max.

In Section \ref{subsec:Intuition}, we  compared excess kurtosis and Max in a  heuristic way;  here we will continue that discussion,  using now empirical results.  Notice that for the data set $(w_1, w_2, \ldots, w_n)$, the 
largest (absolute) observation  is:
\[
M  = M_n =  17.48, 
\]
and the excess kurtosis is:
\[
\kappa = \kappa_n  =  \frac{1}{n} [\sum_i w_i^4]  - 3  = 27.08. 
\]
In the asymptotic analysis of  Section  \ref{subsec:Intuition}, we  
assumed $\kappa(W)$  is a constant.  However for  $n = 244^2$, 
we get a very large 
excess kurtosis $27.08 \approx n^{0.3}$;    this  will make excess kurtosis
very favorable in the current situation.

Now,   in order for $M_n$ to work successfully, we have to 
take $\lam$ to be  large enough that
\[
\sqrt{\lam}  M  > \sqrt{2 \log n}
\]
so  $\lam > 0.072$.  The $p$-value of $M_n$ is then:
\[
  \mathrm{exp}(-\mathrm{exp}(-\frac{\sqrt{\lam}M  - 4.2627}{0.2125})), 
\]
moreover, the $p$-value for excess kurtosis is heuristically
\[
\bar{\Phi}^{-1} (\sqrt{n} \lam^2 \kappa); 
\]
setting them to be equal,  we can solve  $\kappa$ in terms of $M$: 
\[
\kappa = \kappa_0(M).
\]
The curve $\kappa = \kappa_0(M)$ separates the $M$-$\kappa$ plane  into $2$ regions:  the region above the curve  is  favorable to the excess kurtosis, and the region below the curve is  favorable to Max. See Figure \ref{Figure:Region}. 
In the current situation,  the point $(M,\kappa) = (17.48, 27.08)$ falls 
far above the curve;  this explains why excess kurtosis  is  better than Max for 
the current data set. 

%%%%%%%
%%%%%%%
%%%%%%%
{\small
\begin{table}[htb]
\baselineskip=0.4cm
\begin{center}
\begin{tabular}{l|c} \hline \hline
Multi-scale Method           & Alpha         \\ 
\hline \hline
Bi-orthogonal Wavelet        &                 \\
Scale 1, Horizontal          &  6.13           \\
Scale 1, Vertical            &  4.84           \\
Scale 1, Diagonal            &  4.27           \\
Scale 2, Horizontal          &  5.15           \\
Scale 2, Vertical            &  4.19            \\
Scale 2, Diagonal            &  3.83           \\ 
Scale 3, Horizontal          &  4.94           \\
Scale 3, Vertical            &  4.99           \\
Scale 3, Diagonal            &  4.51          \\
Scale 4, Horizontal          &  3.26          \\
Scale 4, Vertical            &  3.37          \\
Scale 4, Diagonal            &  3.76          \\  \hline
\hline
\end{tabular}
\caption{Table of $\alpha$ values for which the different wavelet bands of the CS map.}
\label{tab_cs_alpha}
\end{center}
\end{table}
}

%%%%%%%%%
%%%%%%%%%
%%%%%%%%%
\subsection{Experiments on Wavelet Coefficients}   \label{sec-simulation-experiment}

\subsubsection{CMB + CS}
We study the relative sensitivity of the different wavelet-based statistical methods  when the signals are added to a dominant Gaussian noise, i.e.  the primary CMB.  

We ran 5000 simulations by adding the 100 CMB realisations to the CS
($D(\lambda,i) = \sqrt{1 - \lambda}\mathrm{CMB}_i + \sqrt{\lambda}\mathrm{CS}$,
$i = 1 \dots 100$), using 50 different values for $\lambda$, ranging  linearly between $0$ and $0.18$.
 Then we applied the  bi-orthogonal wavelet transform, using the standard 7/9 filter 
 \cite{wave:antonini92} to these 5000 maps. On  
each band $b$ of the wavelet transform, for each dataset $D(\lambda, i)$, 
we calculate the  kurtosis value ${K}_{D(b, \lambda, i)}$. 
In order to calibrate
and compare the departures from a Gaussian distribution, we have
simulated for each image $D(\lambda,i)$ a Gaussian Random Field
$G(\lambda,i)$ which has the same power spectrum as $D(\lambda, i)$, and we derive 
its kurtosis values ${K}_{G(b, \lambda, i)}$.  
For a given band $b$ and a given $\lambda$, we derive for each each kurtosis ${K}_{D(b, \lambda, i)}$
its p-value $p_K(b,\lambda,i)$ under the null hypothesis (i.e. no CS) using the distribution of 
${K}_{G(b, \lambda, *)}$. The mean p-value ${\bar p}_K(b,\lambda)$ is obtained by taking the 
mean of  $p_K(b,\lambda,*)$.
For a given band $b$, the curve ${\bar p}_K(b,\lambda)$ versus $\lambda$ shows the sensitivity
of the method to detect CS. 
Then we do the same operation, but replacing the kurtosis by HC and Max.
Figure~\ref{fig_wtcmbcs} shows the mean p-value versus $\lambda$ for 
the nine finest scale subbands of the wavelet transform. The first three subbands  correspond  to the finest
scale (high frequencies) in the three directions, respectively horizontal, vertical and diagonal.
Bands 4,5, and 6 correspond  to the second resolution level and bands 7,8,9 to the third.
Results are clearly in favor of the excess kurtosis.

\begin{figure}
\centering
\caption{For the nine first bands of the wavelet transform, the mean p-value versus $\lambda$.
The solid, dashed and dotted lines correspond respectively to the excess kurtosis, the HC and Max. }
\label{fig_wtcmbcs}
\end{figure}

%\begin{figure}[htb]
%\centerline{
%\hbox{
%\psfig{figure=fig_wt_cmb_cs.ps,bbllx=1.5cm,bblly=12.5cm,bburx=19.5cm,bbur%y=25.5cm,height=11cm,width=16.cm,clip=}
%}}
%\caption{For the nine first band of the wavelet transform, the mean p-value %versus $u = 1-\lambda$ is plotted.
%The solid, dashed and dotted lined correspond respectively to the Kurtosis, %the HC and MAX.}
%\label{fig_wtcmbcs}
%\end{figure}

{\small
\begin{table}[htb]
\baselineskip=0.4cm
\begin{center}
\begin{tabular}{l|cccc} \hline \hline
Multi-scale Method           & excess kurtosis&   HC   &  MAX       \\ 
\hline \hline
Bi-orthogonal Wavelet        &          &        &             \\
Scale 1, Horizontal          &   0.73   &  0.73  &   0.73      \\
Scale 1, Vertical            &   0.73   &  0.73  &   0.73      \\
Scale 1, Diagonal            &   0.38   &  0.38  &   0.38      \\
Scale 2, Horizontal          &   8.01   &  9.18  &   8.81      \\
Scale 2, Vertical            &   6.98   &  8.44  &   10.65     \\
Scale 2, Diagonal            &   2.20   &  2.94  &   2.57      \\ \hline
A trous Wavelet Transform    &          &        &             \\
Scale 1                      &  1.47    &   1.47 &   1.47     \\
Scale 2                      &  9.91    & 12.85  &   16.53    \\ \hline
Curvelet                     &          &        &             \\
Scale 1, Band 1              &  1.47    &  2.20  &   3.30      \\
Scale 1, Band 2              &  13.59   &  16.90 &    -        \\
Scale 2, Band 1              &  11.38   &  14.32 &    -         \\ \hline
\hline
\end{tabular}
\caption{Table of $\lambda$ values (multiplied 100) for  
CS detections at 95\% confidence.}
\label{tab_cmb_cs_pval}
\baselineskip=0.4cm
\begin{tabular}{l|cccc} \hline \hline
Multi-scale Method           & excess kurtosis&   HC   &  MAX       \\ 
\hline \hline
Bi-orthogonal Wavelet        &          &        &             \\
Scale 1, Horizontal          &   0.30   &  0.32  &   -        \\
Scale 1, Vertical            &   0.32   &  0.32  &   -        \\
Scale 1, Diagonal            &   0.06   &  0.06  &   0.24      \\
Scale 2, Horizontal          &     -    &   -    &   -      \\
Scale 2, Vertical            &     -    &   -    &   -     \\
Scale 2, Diagonal            &   0.65   &  0.71  &   -      \\ \hline
A trous Wavelet Transform    &          &        &             \\
Scale 1                      &  0.41    &   0.47 &   -     \\  \hline
Curvelet                     &          &        &             \\
Scale 1, Band 1              &  0.59    &  0.69  &   0.83      \\  \hline
\hline
\end{tabular}
\caption{Table of $\lambda$ values  for which the SZ 
detections at  95\% confidence.}
\label{tab_cmb_sz_pval}
\end{center}
\end{table}
}

The same experiments have been repeated, but replacing the bi-orthogonal wavelet transform
by the undecimated isotropic \`a trous wavelet transform. 
Results are similarly in favor of the excess kurtosis. Table~\ref{tab_cmb_cs_pval} gives the $\lambda$ 
values (multiplied 100) for 
which the CS are detected at a 95\% confidence level.
Only bands where this level is achieved are given. Smaller the $\lambda$, better the the sensibility
of the method to the detect the CS. These results show that the excess kurtosis outperforms 
clearly HC and Max, whatever the chosen multiscale transform and the analyzed scale.

No method is able to detect the CS at a 95\% confidence level after the second scale in these
simulations. In practice, the presence of noise makes  the detection even more 
difficult, especially in the finest scales.

%%%%%%%%%%%
%%%%%%%%%%%
%%%%%%%%%%%
\subsubsection{CMB + SZ}
We now consider a totally different contamination.
Here, we take into account the
secondary anisotropies due to the kinetic Sunyaev-Zel'dovich (SZ)
effect \cite{sunyaev80}.  The SZ effect represents the Compton
scattering of  CMB photons by the free electrons of the ionised and
hot intra-cluster gas. When the galaxy cluster moves with respect to
the CMB rest frame, the Doppler shift induces additional anisotropies;
this is the so-called kinetic SZ (KSZ) effect. The kinetic SZ maps are simulated following
Aghanim et al \cite{gauss:aghanim01b} and the simulated observed map $D$ is obtained from
$D_\lambda = \mathrm{CMB}+ \lambda \mathrm{KSZ}$,
where $\mathrm{CMB}$ and $\mathrm{KSZ}$ are
respectively the CMB and the kinetic SZ simulated maps.  
We ran  5000 simulations by adding the 100 CMB realisations to the KSZ
($D(\lambda,i) = \mathrm{CMB}_i +  \lambda \mathrm{KSZ}$,
$i = 1 \dots 100$), using 50 different values for $\lambda$, ranging linearly between 0 and $1$.
The p-values are calculated just as in the previous section.

 Table~\ref{tab_cmb_sz_pval} gives the $\lambda$ values  for 
which SZ is detected at a 95\% confidence level for the three multiscale transforms.
Only bands were this level is achieved are given.
Results are again in favor of the Kurtosis.

\section{Curvelet Coefficients of Filaments}  \label{sec-curve-filaments}
Curvelet Analysis was proposed by Cand\`es and Donoho (1999) \cite{Curvelets-StMalo}
as a means to efficiently represent edges in images; Donoho
and Flesia (2001) \cite{Flesia} showed that it could also be used to describe
non-Gaussian statistics in natural images.  It has also been used
for a variety of image processing tasks:  \cite{CurveInverse,starck:sta01_3,starck:sta02_4}.
We now consider the use of curvelet analysis
for detection of non-Gaussian cosmological structures
which are filamentary.

\subsection{Curvelet Coefficients of Filaments}

Suppose we have an image $I$ which contains within it
a single filament, i.e. a smooth curve of appreciable length $L$.
We analyse it using the curvelet frame.
Applying analysis techniques described carefully
in \cite{CandesDonoho2004}, we can make precise
the following claim: {\sl at scale $s = 2^{-j}$
there will be about $O(L 2^{j/2})$ significant coefficients caused by this
filamentary feature, and they will all be of roughly similar size.
The remaining $O(4^j)$ coefficients at that scale will
be much smaller, basically zero in comparison. }

The pattern continues in this way if there is a collection
of $m$ filaments of individual lengths $L_i$ and total
length $L = L_1 + \dots + L_m$.  Then
we expect roughly $O(L 2^{j/2})$ substantial
coefficients at level $j$, out of $4^j$ total.

This suggests a rough model for the analysis
of non-Gaussian random images which contain apparent `edgelike'
phenomena. If we identify the edges with filaments,
then we expect to see, at a scale with $n$ coefficients,
about $L n^{1/4}$ nonzero coefficients. Assuming
all the edges are equally `pronounced', this suggests that
we view the curvelet coefficients
of $I$ at a given scale as consisting of
a fraction $\eps = L/n^{3/4}$ nonzeros
and the remainder zero.  Under this model,
the curvelet coefficients of a superposition
of a Gaussian random image should behave like:
\begin{equation}     
X_i =  (1 - \eps) N(0,1)  + \frac{\eps}{2}  N(-\mu, 1) +  \frac{\eps}{2}  N(\mu, 1),
\end{equation}
where $\eps$ are the fraction of large curvelet coefficients corresponding
 to filaments, and  $\mu$ is the amplitude of these coefficients of the non-Gaussian component $N$.

 The problem of detecting the existence of such
 a non-Gaussian mixture  is equivalent  to discriminating between the hypotheses:  \begin{align}
&H_0: \;\;\;   X_i \stackrel{iid}{\sim}  N(0,1),  \label{EqHypo3}   \\
&H_1^{(n)}:   X_i = (1 - \eps_n) N(0,1)   + \frac{\eps_n}{2} N(-\mu_n,1) + \frac{\eps_n}{2} N(\mu_n,1),  \label{EqHypo4}
\end{align}
and  $N \equiv 0$ is equivalent to $\eps_n   \equiv  0$.

\subsection{Optimal Detection of Sparse Mixtures}
When both $\eps$ and $\mu$ are known,  the optimal test for Problem (\ref{EqHypo3}) - (\ref{EqHypo4}) is simply the Neyman-Pearson  Likelihood ratio test (LRT), \cite[Page 74]{Lehmann}.  Asymptotic analysis shows the following, \cite{Ingster1999,Jin2004}. 

Suppose we let $\eps_n = n^{-\beta}$ for some exponent
$\beta \in (1/2,1)$, and
\begin{equation} \label{MuDef}
       \mu_n = \sqrt{2s \log(n)},  \quad  0<s<1.
\end{equation}
There is a {\it threshold effect}: setting
\begin{equation}  
\rho^*_2(\beta) =
\left\{ \begin{array}{ll}
 \beta - \frac{1}{2}, &\ \ 1/2 < \beta \leq 3/4,\\
(1-\sqrt{1-\beta})^2, &\ \ 3/4 < \beta < 1.
\end{array}
\right.
\end{equation}
then as $n \goto \infty$,   LRT is able to reliably detect  for large $n$
when  $s > \rho^*_2(\beta)$, and is unable  to detect when $s  <  \rho^*_2(\beta)$, \cite{Jin2004},   \cite{Ingster1999}, and \cite{gauss:lin02}.
Since LRT is optimal,
 it is not possible for any statistic to  reliably detect when $s <  \rho^*_2(\alpha)$.
We call the curve $s = \rho^*_2(\beta)$ in the $\beta$-$s$ plane  the {\it detection boundary}; see Figure \ref{fig:HCDetect}.

%%%%%%%
%%%%%%%
%%%%%%%
\begin{figure}
\centering
\caption{The detection boundary separates the square in the $\beta$-$s$ plane into the detectable region and the undetectable region.  When $(\beta,s)$ falls into the estimable region,  it is possible not only to  reliably detect the presence of the signals, but also estimate them. }
\label{fig:HCDetect}
\end{figure}

We also remark that if the sparsity parameter $\beta < 1/2$,
it is possible to discriminate merely using the value of the empirical variance
of the observations or some other simple moments, and so there is no
need for advanced theoretical approaches.

\subsection{Adaptive Testing using Higher Criticism}  \label{sec:HC}
The Higher Criticism statistic (HC), was  proposed in \cite{gauss:lin02},  where it was proved to be asymptotically optimal in detecting (\ref{EqHypo3}) - (\ref{EqHypo4}).

To define HC  first  we convert
the individual $X_i$'s  into $p$-values for individual
$z$-tests.  Let $p_i = P\{ N(0,1) > X_i \}$ be the $i^{th}$ $p$-value, and let
$p_{(i)}$ denote the $p$-values {\it sorted in increasing order};
the Higher Criticism statistic  is defined as:
\[
       HC_{n}^* =  \max_{i}
         \biggl| \sqrt{n} [i/n  - p_{(i)}]/ \sqrt{p_{(i)} (1-p_{(i)})} \biggr|,
\]
or in a modified form:
\[
HC_n^+  = \max_{\{i:  \; 1/n  \leq  p_{(i)} \leq  1 - 1/n \}}
         \biggl|   \sqrt{n} [i/n  - p_{(i)}]/ \sqrt{p_{(i)} (1-p_{(i)})}  \biggr|;
\]
we let  $HC_n$ refer either to  $HC_n^*$  or $HC_n^+$ whenever there is no confusion.
The above definition is slightly different from \cite{gauss:lin02}, but  the ideas are essentially the same.

With an appropriate normalization sequence:
\[
a_n = \sqrt{2 \log \log n}, \qquad b_n = 2 \log \log n + 0.5 \log \log \log n - 0.5 \log (4 \pi),
\]
the distribution of $HC_n$ converges  to  the Gumbel distribution $E_v^4$,
whose cdf is $\mathrm{exp}(-4\mathrm{exp}(-x))$, \cite{Shorack}:
\[
a_n  HC_n  - b_n  \rightarrow_w  E_v^4,
\]
so  the $p$-values of $HC_n$ are approximately:
\begin{equation}  \label{EqHCP}
\mathrm{exp}(-4\mathrm{exp}( - [a_n HC_n - b_n])).
\end{equation}
For moderately large $n$,  in general,  the approximation in (\ref{EqHCP})
is accurate for the $HC_n^+$, but not for
$HC_n^*$.  For  $n = 244^2$,  taking $a_n = 2.2536 $ and
$b_n = 3.9407$ in (\ref{EqHCP}) gives a good
approximation for the $p$-value of $HC_n^+$.

A brief remark comparing Max and  HC.
Max only takes into account  the few largest observations,
HC takes into  account those outliers,  but also
moderate large observations;  as a result, in general,
HC is  better than Max, especially when
we have unusually many moderately large observations.
However,  when the actual evidence lies in the middle of the
distribution  both HC and Max will be very weak.

%%%%%%%%%
%%%%%%%%%
%%%%%%%%%
\subsection{Curvelet Coefficients of Cosmic Strings}
In Section \ref{sec:AlgTail},  we  studied wavelet coefficients of  simulated cosmic strings. We now study the curvelet coefficients of the same simulated maps.

%%%%%%%
%%%%%%%
\begin{figure}
\centering
\caption{From top to bottom: the image of the bar, the log-histogram of the curvelet coefficients of the bar, and the qqplot of the curvelet coefficients vs. Normal.}
\label{Figure:ArtCur}
\end{figure}

We now discuss  empirical properties of Curvelet coefficients
of (simulated) cosmic strings.  
% We used a recently developed
% digital curvelet transform of Cand\`es and Donoho (Manuscript in
% preparation).
This was first deployed on a test image showing a simple
`bar' extending vertically across the image. The result,
seen in Figure \ref{Figure:ArtCur} shows the image,
the histogram of the curvelet coefficients at the
next-to-finest scale, and the qq-plot against
the normal distribution.  The display matches in general
terms the sparsity model of section \ref{sec-curve-filaments}.
That display also shows the result of superposing
Gaussian noise on the image; the curvelet coefficients clearly
have the general appearance of a mixture of normals
with sparse fractions at nonzero mean, just as in the model.

We also applied the curvelet transform to the
simulated cosmic string data. Figure \ref{Figure:CsCur}
shows the results, which suggest that
the coefficients do not match the simple sparsity model.
Extensive modelling efforts, not reported
here, show that  the curvelet 
coefficients  transformed by $|v|^{0.815}$ have
 an exponential distribution.

This discrepancy from the sparsity model
has two explanations.  First, cosmic string
images contain (to the naked eye) both point-like
features and curvelike features.  Because curvelets
are not specially adapted to sparsifying point-like features,
the coefficients contain extra information not expressible
by our geometric model.  Second, cosmic string
images might contain filamentary features at a
range of length scales and a range of density contrasts.
If those contrasts exhibit substantial amplitude variation,
the simple mixture model must be replaced by something more
complex.  In any event, the curvelet coefficients
of cosmic strings do not have the simple structure
proposed in Section \ref{sec-curve-filaments}.

When applying various detectors
of non-Gaussian behavior to curvelet coefficients,
as in the simulation of Section  \ref{sec-simulation-experiment},
we find that, despite the theoretical ideas backing the use
of HC as an optimal test for sparse non-Gaussian phenomena,
the kurtosis consistently has better performance.
The results are included in Tables \ref{tab_cmb_cs_pval} 
 and  \ref{tab_cmb_sz_pval}.

Note that, although the curvelet coefficients
are not as sensitive detectors as wavelets
in this setting, that can be an advantage, since they
are relatively immune to point-like features
such as SZ contaimination. Hence they are more
specific to CS as opposed to SZ effects.

%%%%%%%
%%%%%%%
\begin{figure}
\centering
\caption{For the curvelet coefficients $v_i$'s of the simulated CS map 
in Figure \ref{fig_cmb},  from top to bottom: the log-histogram of $v_i$'s,  
the qqplot of $v_i$'s  vs. Normal, and the qqplot of $\mathrm{sign}(v_i) |v_i|^{0.815}$  vs.  double exponential.}
\label{Figure:CsCur}
\end{figure}

\section{Conclusion}
The kurtosis of the wavelet coefficients is very often used in astronomy for the 
detection non-Gaussianities in the CMB.  It has been shown \cite{starck:sta03_1} that it is also
possible to separate the non-Gaussian signatures associated with
 cosmic strings from those due to SZ effect by combining the excess 
kurtosis derived from these both the curvelet and the wavelet transform.  
We have studied in this paper several other transform-based statistics, the MAX and the 
Higher Criticism, and we have compared them theoretically and experimentally to the kurtosis.
We have shown that kurtosis is asymptotically optimal in the class of weakly dependent symmetric
non-Gaussian contamination with finite 8-th moments, while HC and MAX are asymptotically optimal in the class 
of weakly dependent symmetric non-Gaussian contamination with infinite 8-th moment.
Hence depending on the nature of the non-Gaussianity, a statitic is better than another one. This is a motivation
for using several statistics rather than a single one, for analysing  CMB data. 
Finally, we have studied in details the case of cosmic string contaminations on simulated maps.
Our experiment results show clearly that kurtosis outperforms Max/HC.

\section*{Acknowledgments}
The cosmic string maps were kindly provided by F.R. Bouchet.  The authors would also like to thank Inam Rahman for help in simulations. 

\vspace{1em}% 
\bibliographystyle{plain}
% \bibliographystyle{alpha}
% \bibliography{gauss,candes,entropy,starck,wave,restore,ima,astro,compress,mc,curvelet,nab}

\begin{thebibliography}{10}

\bibitem{gauss:aghanim99}
N.~{Aghanim} and O.~{Forni}.
\newblock {Searching for the non-Gaussian signature of the CMB secondary
  anisotropies}.
\newblock {\em Astronomy and Astrophysics}, 347:409--418, July 1999.

\bibitem{gauss:aghanim01b}
N.~{Aghanim}, K.~M. {G{\' o}rski}, and J.-L. {Puget}.
\newblock {How accurately can the SZ effect measure peculiar cluster velocities
  and bulk flows?}
\newblock {\em Astronomy and Astrophysics}, 374:1--12, July 2001.

\bibitem{gauss:aghanim03}
N.~{Aghanim}, M.~{Kunz}, P.~G. {Castro}, and O.~{Forni}.
\newblock {Non-Gaussianity: Comparing wavelet and Fourier based methods}.
\newblock {\em Astronomy and Astrophysics}, 406:797--816, August 2003.

\bibitem{wave:antonini92}
M.~Antonini, M.~Barlaud, P.~Mathieu, and I.~Daubechies.
\newblock Image coding using wavelet transform.
\newblock {\em IEEE Transactions on Image Processing}, 1:205--220, 1992.

\bibitem{gauss:barreiro01}
R.~B. {Barreiro} and M.~P. {Hobson}.
\newblock {The discriminating power of wavelets to detect non-Gaussianity in
  the cosmic microwave background}.
\newblock {\em Monthly Notices of the Royal Astronomical Society},
  327:813--828, November 2001.

\bibitem{astro:bennett2003}
C.~L. {Bennett}, M.~{Halpern}, G.~{Hinshaw}, N.~{Jarosik}, A.~{Kogut},
  M.~{Limon}, S.~S. {Meyer}, L.~{Page}, D.~N. {Spergel}, G.~S. {Tucker},
  E.~{Wollack}, E.~L. {Wright}, C.~{Barnes}, M.~R. {Greason}, R.~S. {Hill},
  E.~{Komatsu}, M.~R. {Nolta}, N.~{Odegard}, H.~V. {Peiris}, L.~{Verde}, and
  J.~L. {Weiland}.
\newblock {First-Year Wilkinson Microwave Anisotropy Probe (WMAP) Observations:
  Preliminary Maps and Basic Results}.
\newblock {\em Astrophysical Journal, Supplement Series}, 148:1--27, September
  2003.

\bibitem{gauss:bennett03}
C.~L. {Bennett}, R.~S. {Hill}, G.~{Hinshaw}, M.~R. {Nolta}, N.~{Odegard},
  L.~{Page}, D.~N. {Spergel}, J.~L. {Weiland}, E.~L. {Wright}, M.~{Halpern},
  N.~{Jarosik}, A.~{Kogut}, M.~{Limon}, S.~S. {Meyer}, G.~S. {Tucker}, and
  E.~{Wollack}.
\newblock {First-Year Wilkinson Microwave Anisotropy Probe (WMAP) Observations:
  Foreground Emission}.
\newblock {\em Astrophysical Journal, Supplement Series}, 148:97--117,
  September 2003.

\bibitem{gauss:benoit03}
A.~{Beno{\^ i}t}, P.~{Ade}, A.~{Amblard}, R.~{Ansari}, {\' E}.~{Aubourg},
  S.~{Bargot}, J.~G. {Bartlett}, J.-P. {Bernard}, R.~S. {Bhatia},
  A.~{Blanchard}, J.~J. {Bock}, A.~{Boscaleri}, F.~R. {Bouchet},
  A.~{Bourrachot}, P.~{Camus}, F.~{Couchot}, P.~{de Bernardis},
  J.~{Delabrouille}, F.-X. {D{\' e}sert}, O.~{Dor{\' e}}, M.~{Douspis},
  L.~{Dumoulin}, X.~{Dupac}, P.~{Filliatre}, P.~{Fosalba}, K.~{Ganga},
  F.~{Gannaway}, B.~{Gautier}, M.~{Giard}, Y.~{Giraud-H{\' e}raud},
  R.~{Gispert}, L.~{Guglielmi}, J.-C. {Hamilton}, S.~{Hanany},
  S.~{Henrot-Versill{\' e}}, J.~{Kaplan}, G.~{Lagache}, J.-M. {Lamarre}, A.~E.
  {Lange}, J.~F. {Mac{\'{\i}}as-P{\' e}rez}, K.~{Madet}, B.~{Maffei},
  C.~{Magneville}, D.~P. {Marrone}, S.~{Masi}, F.~{Mayet}, A.~{Murphy},
  F.~{Naraghi}, F.~{Nati}, G.~{Patanchon}, G.~{Perrin}, M.~{Piat},
  N.~{Ponthieu}, S.~{Prunet}, J.-L. {Puget}, C.~{Renault}, C.~{Rosset},
  D.~{Santos}, A.~{Starobinsky}, I.~{Strukov}, R.~V. {Sudiwala}, R.~{Teyssier},
  M.~{Tristram}, C.~{Tucker}, J.-C. {Vanel}, D.~{Vibert}, E.~{Wakui}, and
  D.~{Yvon}.
\newblock {The cosmic microwave background anisotropy power spectrum measured
  by Archeops}.
\newblock {\em Astronomy and Astrophysics}, 399:L19--L23, March 2003.

\bibitem{bernardeau2002}
F.~{Bernardeau} and J.~{Uzan}.
\newblock {Non-Gaussianity in multifield inflation}.
\newblock {\em Phys. Rev. D}, 66:103506--+, November 2002.

\bibitem{gauss:bouchet00}
F.~R. {Bouchet}, P.~{Peter}, A.~{Riazuelo}, and M.~{Sakellariadou}.
\newblock {Evidence against or for topological defects in the BOOMERanG data?}
\newblock {\em Phys. Rev. D}, 65:21301--+, January 2002.

\bibitem{gauss:bromley99}
B.~C. {Bromley} and M.~{Tegmark}.
\newblock {Is the Cosmic Microwave Background Really Non-Gaussian?}
\newblock {\em Astrophysical Journal Letter}, 524:L79--L82, October 1999.

\bibitem{Curvelets-StMalo}
E.~J. Cand\`es and D.~L. Donoho.
\newblock Curvelets -- a surprisingly effective nonadaptive representation for
  objects with edges.
\newblock In A.~Cohen, C.~Rabut, and L.L. Schumaker, editors, {\em Curve and
  Surface Fitting: Saint-Malo 1999}, Nashville, TN, 1999. Vanderbilt University
  Press.

\bibitem{CurveInverse}
E.~J. Cand\`es and D.~L. Donoho.
\newblock Edge-preserving denoising in linear inverse problems: optimality of
  curvelet frames.
\newblock {\em Ann. Statist}, 30(3):784--842, 2002.

\bibitem{CandesDonoho2004}
E.~J. Cand\`es and D.~L. Donoho.
\newblock frames of curvelets and optimal representations of objects with
  piecewise $c\sp 2$ singularities.
\newblock {\em Comm. Pure Appl. Math}, 57(2):219--266, 2004.

\bibitem{gauss:cayon01}
L.~{Cay{\' o}n}, J.~L. {Sanz}, E.~{Mart{\' i}nez-Gonz{\' a}lez}, A.~J.
  {Banday}, F.~{Arg{\" u}eso}, J.~E. {Gallegos}, K.~M. {G{\' o}rski}, and
  G.~{Hinshaw}.
\newblock {Spherical Mexican hat wavelet: an application to detect
  non-Gaussianity in the COBE-DMR maps}.
\newblock {\em Monthly Notices of the Royal Astronomical Society},
  326:1243--1248, October 2001.

\bibitem{gauss:bernardis00}
P.~{de Bernardis}, P.~A.~R. {Ade}, J.~J. {Bock}, J.~R. {Bond}, J.~{Borrill},
  A.~{Boscaleri}, K.~{Coble}, B.~P. {Crill}, G.~{De Gasperis}, P.~C. {Farese},
  P.~G. {Ferreira}, K.~{Ganga}, M.~{Giacometti}, E.~{Hivon}, V.~V. {Hristov},
  A.~{Iacoangeli}, A.~H. {Jaffe}, A.~E. {Lange}, L.~{Martinis}, S.~{Masi},
  P.~V. {Mason}, P.~D. {Mauskopf}, A.~{Melchiorri}, L.~{Miglio}, T.~{Montroy},
  C.~B. {Netterfield}, E.~{Pascale}, F.~{Piacentini}, D.~{Pogosyan},
  S.~{Prunet}, S.~{Rao}, G.~{Romeo}, J.~E. {Ruhl}, F.~{Scaramuzzi},
  D.~{Sforna}, and N.~{Vittorio}.
\newblock {A flat Universe from high-resolution maps of the cosmic microwave
  background radiation}.
\newblock {\em Nature}, 404:955--959, April 2000.

\bibitem{Flesia}
D.~L. Donoho and A.G. Flesia.
\newblock Can recent developments in harmonic analysis explain the recent
  findings in natural scene statistics?
\newblock {\em Network: Computation in Neural Systems}, 12(3):371--393, 2001.

\bibitem{DJ04b}
D.~L. Donoho and J.~Jin.
\newblock Optimality of excess kurtosis for detecting a non-gaussian component
  in high-dimensional random vectors.
\newblock Technical report, Stanford University, 2004.

\bibitem{gauss:lin02}
D.L. Donoho and J.~Jin.
\newblock Higher criticism for detecting sparse heterogeneous mixtures.
\newblock {\em Ann. Statist.}, 32(3):962-- 994, 2004.

\bibitem{gauss:fixsen96}
D.~J. {Fixsen}, E.~S. {Cheng}, D.~A. {Cottingham}, W.~C. {Folz}, C.~A. {Inman},
  M.~S. {Kowitt}, S.~S. {Meyer}, L.~A. {Page}, J.~L. {Puchalla}, J.~E. {Ruhl},
  and R.~F. {Silverberg}.
\newblock {A Balloon-borne Millimeter-Wave Telescope for Cosmic Microwave
  Background Anisotropy Measurements}.
\newblock {\em Astrophysical Journal}, 470:63--+, October 1996.

\bibitem{gauss:forni99}
O.~{Forni} and N.~{Aghanim}.
\newblock {Searching for non-gaussianity: Statistical tests}.
\newblock {\em Astronomy and Astrophysics, Supplement Series}, 137:553--567,
  June 1999.

\bibitem{gauss:gangui94}
A.~{Gangui}, F.~{Lucchin}, S.~{Matarrese}, and S.~{Mollerach}.
\newblock {The three-point correlation function of the cosmic microwave
  background in inflationary models}.
\newblock {\em Astrophysical Journal}, 430:447--457, August 1994.

\bibitem{gauss:halverson02}
N.~W. {Halverson}, E.~M. {Leitch}, C.~{Pryke}, J.~{Kovac}, J.~E. {Carlstrom},
  W.~L. {Holzapfel}, M.~{Dragovan}, J.~K. {Cartwright}, B.~S. {Mason},
  S.~{Padin}, T.~J. {Pearson}, A.~C.~S. {Readhead}, and M.~C. {Shepherd}.
\newblock {Degree Angular Scale Interferometer First Results: A Measurement of
  the Cosmic Microwave Background Angular Power Spectrum}.
\newblock {\em Astrophysical Journal}, 568:38--45, March 2002.

\bibitem{gauss:hanany00}
S.~{Hanany}, P.~{Ade}, A.~{Balbi}, J.~{Bock}, J.~{Borrill}, A.~{Boscaleri},
  P.~{de Bernardis}, P.~G. {Ferreira}, V.~V. {Hristov}, A.~H. {Jaffe}, A.~E.
  {Lange}, A.~T. {Lee}, P.~D. {Mauskopf}, C.~B. {Netterfield}, S.~{Oh},
  E.~{Pascale}, B.~{Rabii}, P.~L. {Richards}, G.~F. {Smoot}, R.~{Stompor},
  C.~D. {Winant}, and J.~H.~P. {Wu}.
\newblock {MAXIMA-1: A Measurement of the Cosmic Microwave Background
  Anisotropy on Angular Scales of 10'-5 degrees}.
\newblock {\em Astrophysical Journal Letter}, 545:L5--L9, December 2000.

\bibitem{gauss:hill75}
B.M. Hill.
\newblock A simple general approach to inference about the tail of a
  distribution.
\newblock {\em Ann. Statist.}, 3:1163--1174, 1975.

\bibitem{gauss:hobson99}
M.~P. {Hobson}, A.~W. {Jones}, and A.~N. {Lasenby}.
\newblock {Wavelet analysis and the detection of non-Gaussianity in the cosmic
  microwave background}.
\newblock {\em Monthly Notices of the Royal Astronomical Society},
  309:125--140, October 1999.

\bibitem{Ingster1999}
Y.~I. Ingster.
\newblock Minimax detection of a signal for $l^p_n$-balls.
\newblock {\em Math. Methods Statist}, 7:401--428, 1999.

\bibitem{gauss:jaffe94}
A.~H. {Jaffe}.
\newblock {Quasilinear evolution of compensated cosmological perturbations: The
  nonlinear {$\sigma$} model}.
\newblock {\em Phys. Rev. D}, 49:3893--3909, April 1994.

\bibitem{gauss:jewell01}
J.~{Jewell}.
\newblock {A Statistical Characterization of Galactic Dust Emission as a
  Non-Gaussian Foreground of the Cosmic Microwave Background}.
\newblock {\em Astrophysical Journal}, 557:700--713, August 2001.

\bibitem{Jin2004}
J.~Jin.
\newblock Detecting a target in very noisy data from multiple looks.
\newblock {\em {IMS} Lecture Notes Monograph}, 45:1--32, 2004.

\bibitem{gauss:kunz01}
M.~{Kunz}, A.~J. {Banday}, P.~G. {Castro}, P.~G. {Ferreira}, and K.~M. {G{\'
  o}rski}.
\newblock {The Trispectrum of the 4 Year COBE DMR Data}.
\newblock {\em Astrophysical Journal Letter}, 563:L99--L102, December 2001.

\bibitem{Lehmann}
E.~L. Lehmann.
\newblock {\em Testing Statistical Hypotheses, 2nd ed.}
\newblock John Wiley \& Sons, 1986.

\bibitem{gauss:luo94}
X.~{Luo}.
\newblock {The angular bispectrum of the cosmic microwave background}.
\newblock {\em Astrophysical Journal Letter}, 427:L71--L74, June 1994.

\bibitem{ima:mallat98}
S.~Mallat.
\newblock {\em A Wavelet Tour of Signal Processing}.
\newblock Academic Press, 1998.

\bibitem{ROC}
C.E. Metz.
\newblock {\em Basic principles of {ROC} analysis}.
\newblock Semin Nuclear Med, 1978.

\bibitem{gauss:miller99}
A.~D. {Miller}, R.~{Caldwell}, M.~J. {Devlin}, W.~B. {Dorwart}, T.~{Herbig},
  M.~R. {Nolta}, L.~A. {Page}, J.~{Puchalla}, E.~{Torbet}, and H.~T. {Tran}.
\newblock {A Measurement of the Angular Power Spectrum of the Cosmic Microwave
  Background from L = 100 to 400}.
\newblock {\em Astrophysical Journal Letter}, 524:L1--L4, October 1999.

\bibitem{gauss:novikov00}
D.~{Novikov}, J.~{Schmalzing}, and V.~F. {Mukhanov}.
\newblock {On non-Gaussianity in the cosmic microwave background}.
\newblock {\em Astronomy and Astrophysics}, 364:17--25, December 2000.

\bibitem{ostriker86}
J.~P. {Ostriker} and E.~T. {Vishniac}.
\newblock {Generation of microwave background fluctuations from nonlinear
  perturbations at the ERA of galaxy formation}.
\newblock {\em Astrophysical Journal}, 306:L51--L54, July 1986.

\bibitem{gauss:penzias65}
A.~A. {Penzias} and R.~W. {Wilson}.
\newblock {Measurement of the Flux Density of CAS a at 4080 Mc/s.}
\newblock {\em Astrophysical Journal}, 142:1149--+, October 1965.

\bibitem{gauss:phillips01}
N.~G. {Phillips} and A.~{Kogut}.
\newblock {Statistical Power, the Bispectrum, and the Search for
  Non-Gaussianity in the Cosmic Microwave Background Anisotropy}.
\newblock {\em Astrophysical Journal}, 548:540--549, February 2001.

\bibitem{gauss:shandarin02}
S.~F. {Shandarin}.
\newblock {Testing non-Gaussianity in cosmic microwave background maps by
  morphological statistics}.
\newblock {\em Monthly Notices of the Royal Astronomical Society}, 331:865+,
  April 2002.

\bibitem{Shorack}
G.~R. Shorack and J.~A. Wellner.
\newblock {\em Empirical Processes with Applications to Statistics}.
\newblock John Wiley \& Sons, 1986.

\bibitem{gauss:smoot92}
G.~F. {Smoot}, C.~L. {Bennett}, A.~{Kogut}, E.~L. {Wright}, J.~{Aymon}, N.~W.
  {Boggess}, E.~S. {Cheng}, G.~{de Amici}, S.~{Gulkis}, M.~G. {Hauser},
  G.~{Hinshaw}, P.~D. {Jackson}, M.~{Janssen}, E.~{Kaita}, T.~{Kelsall},
  P.~{Keegstra}, C.~{Lineweaver}, K.~{Loewenstein}, P.~{Lubin}, J.~{Mather},
  S.~S. {Meyer}, S.~H. {Moseley}, T.~{Murdock}, L.~{Rokke}, R.~F. {Silverberg},
  L.~{Tenorio}, R.~{Weiss}, and D.~T. {Wilkinson}.
\newblock {Structure in the COBE differential microwave radiometer first-year
  maps}.
\newblock {\em Astrophysical Journal Letter}, 396:L1--L5, September 1992.

\bibitem{starck:sta03_1}
J.-L. {Starck}, N.~{Aghanim}, and O.~{Forni}.
\newblock {Detection and discrimination of cosmological non-Gaussian signatures
  by multi-scale methods}.
\newblock {\em Astronomy and Astrophysics}, 416:9--17, March 2004.

\bibitem{starck:sta01_3}
J.-L. Starck, E.~Cand\`es, and D.L. Donoho.
\newblock The curvelet transform for image denoising.
\newblock {\em IEEE Transactions on Image Processing}, 11(6):131--141, 2002.

\bibitem{starck:book98}
J.-L. Starck, F.~Murtagh, and A.~Bijaoui.
\newblock {\em Image Processing and Data Analysis: The Multiscale Approach}.
\newblock Cambridge University Press, 1998.

\bibitem{starck:sta02_4}
J.-L. Starck, F.~Murtagh, E.~Candes, and D.L. Donoho.
\newblock Gray and color image contrast enhancement by the curvelet transform.
\newblock {\em IEEE Transactions on Image Processing}, 12(6):706--717, 2003.

\bibitem{sunyaev80}
R.~A. {Sunyaev} and I.~B. {Zeldovich}.
\newblock {Microwave background radiation as a probe of the contemporary
  structure and history of the universe}.
\newblock {\em Annual review of astronomy and astrophysics}, 18:537--560, 1980.

\bibitem{gauss:verde00}
L.~{Verde}, L.~{Wang}, A.~F. {Heavens}, and M.~{Kamionkowski}.
\newblock {Large-scale structure, the cosmic microwave background and
  primordial non-Gaussianity}.
\newblock {\em Monthly Notices of the Royal Astronomical Society},
  313:141--147, March 2000.

\bibitem{vishniac87}
E.~T. {Vishniac}.
\newblock {Reionization and small-scale fluctuations in the microwave
  background}.
\newblock {\em Astrophysical Journal}, 322:597--604, November 1987.

\bibitem{white2002}
M.~{White} and J.~D. {Cohn}.
\newblock The theory of anisotropies in the cosmic microwave background.
\newblock astro-ph/0203120, 2002.

\end{thebibliography}

%%%%%%%%
%%%%%%%%
\end{document}